\documentstyle[latexsym,psfig]{mn}

\def\degr{\hbox{$^\circ$}}

\def\farcm{\hbox{$.\mkern-4mu^\prime$}}
\def\farcs{\hbox{$.\!\!^{\prime\prime}$}}               
\def\sqd{\hbox{$^{\Box}$}}
\def\fsqd{\hbox{$.\!\!^{\Box}$}}

\begin{document} 
\title[The Faint Sky Variability Survey]{The Faint Sky Variability
Survey I: An Overview}
\author[P.~J.~Groot et al.]{P.J. Groot$^1$, P.M. Vreeswijk$^2$,
M.E. Everett$^3$, S.B. Howell$^3$, J. van Paradijs$^{2,4;\dagger}$ 
\and M.E. Huber$^3$, E.P.J. van den Heuvel$^2$, T. Augusteijn$^5$,
H. B\"ohnhardt$^6$, P.A. Charles$^7$, 
\and D. Davis$^8$, T.J. Galama$^9$, E. Kuulkers$^{10,11}$, C. Kouveliotou$^{12}$,C. Moreno$^{13}$
\and C. Neese$^{8}$, G. Nelemans$^2$, R. Rebolo$^{14}$, R.G.M. Rutten$^5$, H. Scholl$^{15}$, J. Storm$^{16}$, 
\and N. Tanvir$^{17}$, L.B.F.M. Waters$^2$, R.A.M.J. Wijers$^{18}$\\
$^1$Harvard-Smithsonian Center for Astrophysics, 60 Garden Street,
Cambridge, 02138 MA, USA\\
$^2$Astronomical Institute `Anton Pannekoek'/ CHEAF, Kruislaan
403, 1098 SJ, Amsterdam, The Netherlands\\
$^3$ Astrophysics Group, Planetary Science Institute, 620 N. 6th Ave., Tucson, AZ,  USA\\
$^4$ Physics Department, University of Alabama in Huntsville,
Huntsville, USA\\
$^{5}$ Isaac Newton Group of Telescopes, Apartado de Correos 321, Sta Cruz
de La Palma, La Palma, Spain\\
$^6$ European Southern Observatory, Casilla 19001, Santiago 19, Chile\\
$^7$ Astronomy Department, University of Southampton, Southampton, UK\\
$^8$ Planetary Science Institute, 620 N. 6th Ave., Tucson, AZ,  USA\\
$^9$ Astronomy Department, Californian Institute of Technology,
Pasadena, CA, USA\\ 
$^10$ Space Research Organisation Netherlands, Sorbonnelaan 2,
3584 CA, Utrecht, The Netherlands\\
$^{11}$ Astronomical Institute, Utrecht University, P.O. Box 80.000,
3507 TA Utrecht, The Netherlands\\
$^{12}$ USRA at Marshall Space Flight Center, NASA, Huntsville, USA\\
$^{13}$ Nordic Optical Telescope, La Palma, Spain\\
$^{14}$ Inst\'{\i}tuto de Astrof\'{\i}sica de Canarias, La Laguna,
Tenerife, Spain\\
$^{15}$ Observatoire de la C\^ote d'Azur, Nice, France\\
$^{16}$ Astrophysikalische Institut Postdam, An der Sternwarte 16,
14482 Potsdam, Germany\\
$^{17}$ Astronomy Department, University of Hertfordshire,
Hertfordshire, UK\\
$^{18}$ Department of Physics and Astronomy, SUNY, Stony Brook, NY
11794-3800, USA}

\maketitle

\begin{abstract}
The Faint Sky Variability Survey is aimed at finding variable objects
in the brightness range between 17th and 25th magnitude on timescales
between tens of minutes and years with photometric precisions ranging
from 3 millimagnitudes for the brightest to 0.2 magnitudes for the
faintest objects. An area of at least 50 square degrees, located at
mid-galactic latitudes, will be covered using the Wide Field Camera on
the 2.5m Isaac Newton Telescope on La Palma. The survey
started in November 1998 as part of the INT Wide Field Survey
program. Here we describe the main goals of the Faint Sky Variability
Survey, the methods used in extracting the relevant information and
the future prospects of the survey.
\end{abstract}

\section{Introduction \label{sec:intro}}
The advance of large format ($>$2k$\times$2k) CCDs with high quantum
efficiency has opened up a new area in Galactic and extragalactic 
astrophysics: the systematic study of astrophysical objects fainter
than 20th magnitude. The importance of this brightness regime is
nicely illustrated by the current, fast development in the field of
Gamma-ray Bursts (GRBs; for recent reviews see Katz, Piran and Sari,
1998; Piran, 1999; and Van Paradijs, Kouveliotou and Wijers, 2000), 
where the localization of faint 
variable optical counterparts has led to a large increase in 
our understanding of gamma-ray bursts. 

The Faint Sky Variability Survey
(FSVS\footnote{http://www.astro.uva.nl/$\sim$fsvs} ) started in
November 1998. This survey is aimed at finding variable objects in the
brightness range between 17th and 25th magnitude on timescales between
tens of minutes and years.  In the following sections we will outline
the main goals of the survey (Sect.\,2), the INT Wide Field Camera
(Sect.\,3), the observing strategy (Sect.\,4) and field selection
(Sect.\,5).  After a short comparison with other, running surveys
(Sect.\,6), we will discuss ata reduction (Sect.\,7), final
data products (Sect.\,8), availability of the data (Sect.\,9) and
a short overview of the current status of the Survey (Sect.\,10). An
overview of the general results of the first year of observations will
be given in an accompanying paper (Everett et al., 2000).

\section{Goals of the FSVS \label{sec:goals}}

Understanding the variability of stars 
has often been crucial in the development of
astrophysics, with applications ranging from the evolution of stars, to
the structure of our Galaxy and the distance scale of the Universe. 
Variability studies are currently mainly restricted to either bright
regimes (brighter than 20th magnitude) or very small areas (supernovae
and GRB searches). In the galactic realm, a deep variability study
will not only reveal the characteristics of specific groups of 
stellar objects, but will also shed light on the outer parts of our 
Solar System, the direct Solar Neighbourhood, the structure of our Galaxy,
and the extent of the Galactic Halo. The FSVS is aimed at observing 
at least 50 square degrees (50\sqd) down to 25th magnitude. 
The main targets can be divided into two broad areas of 
interest: photometrically and astrometrically
variable objects.

\subsection{Photometrically variable objects \label{sec:photgoals}}
Among the various classes of variables stars our main targets are:\\
{\bf $\bullet$ Close Binaries}: 
Current detections of low-mass close-binary systems (Cataclysmic
Variables, Low-mass x-ray binaries (including Soft X-ray Transients) 
and AM CVn stars) are strongly biased to small subsets of their populations. Of
these systems the Cataclysmic Variables (CVs) form the main subgroup we
expect to find. We refer to Warner (1995) for an extensive review
of CV properties. 
Currently, most CVs are either found as by-products of
extragalactic studies like blue-excess, quasar surveys (e.g. the Palomar-Green
survey: Green, Schmidt and Liebert, 1986; the Hamburg(/ESO) Quasar
Survey: Engels et al., 1994; Wisotzki et al., 1996; and the
Edingburgh-Cape Survey: Stobie et al., 1988), 
or by their outbursts in which the system suddenly brightens 3-10
magnitudes due to an instability in the accretion disk. However, theoretical
calculations show that the majority of the CV population should have evolved
down to mass-transfer rates that are lower than $\sim$10$^{-11}$
M$_{\odot}$\,yr$^{-1}$ (see e.g. Kolb 1993; Howell, Rappaport and
Politano 1996; Howell, Nelson and Rappaport, 2000).
At these very low-mass transfer
rates CVs are expected to be faint (typically V$>$20), 
have no UV excess, show no (frequent) outbursts,
and will therefore not show up in
conventional searches. However, all CVs show intrinsic variability of the
order of tenths of magnitudes or more. This variability is either caused by
`flickering' (mass-transfer instabilities), 
orbital modulations (hot-spots or eclipses) or long-term
mass-transfer fluctuations. Searching for faint variable stars is
therefore a very good way to disclose the characteristics of the majority
of the CV population. The same search technique will also make the survey
sensitive to other classes of close binaries, such as low-mass x-ray
binaries, soft X-ray transients in quiescence and AM CVn
stars. Since their space densities are much lower than that
of CVs, we expect fewer of these in our Survey. However, because a much 
smaller number of these systems is known, even the discovery of a
few can be a major contribution to the field.\\
{\bf $\bullet$ RR Lyrae:}
Due to their standard candle properties and easy recognition by 
colour and variability, RR Lyrae stars can be used as excellent
tracers of the structure of the galactic halo. A few of these stars
have been found at large galactocentric distances (Hawkins, 1984;
Ciardullo et al., 1989), but number statistics are still poor. Finding more
of these stars will help to constrain the total enveloped mass in the
Galaxy at different radii. \\
{\bf $\bullet$ Optical Transients to Gamma-Ray Bursts}
The detection of optical counterparts to $\gamma$-ray bursts (GRBs,
e.g. Van Paradijs et al., 1997), and the subsequent classification of
GRBs as cosmological (e.g. Metzger et al., 1997, Kulkarni et al.,
1998) have shown that GRBs are among the most energetic phenomena
known in the Universe. The high energies implied by observations of
GRB afterglows (10$^{53-54}$ erg in $\gamma$-rays if isotropy is assumed, 
Kulkarni et al., 1998; 1999), raises the question whether GRBs are emitting their
energy isotropically or in the form of jets. In the latter case the
energies involved will be much lower, depending on the amount of
beaming. Even if the $\gamma$-rays are beamed the optical afterglow is
expected to radiate more isotropically, and thus one expects to observe faint
afterglows without an accompanying burst in $\gamma$-rays. Detections
or non-detections of such transient events will constrain the beaming
angle. A discussion and analysis of such results will be presented 
in Vreeswijk et al. (2001).

\subsection{Astrometrically variable objects \label{sec:astgoals}}
The observing schedule that we have adopted for the FSVS (see 
Sect.\,4) also allows for the detection of
astrometrically variable objects. Our interests fall into two main categories:\\
{\bf $\bullet$ Kuiper Belt Objects}: Kuiper
Belt Objects (KBOs) are icy bodies revolving around the Sun in orbits that
lie outside the orbit of Neptune (which has led to the alternative
name of Trans Neptunian Objects; TNOs). Since their discovery in 1993
(Jewitt and Luu, 1993), more than 100 of these objects have been
found. Studying their properties will give important insight into the
formation of the Solar system and planetary systems in general. 
One question that is particularly well suited to be answered is the
inclination distribution of KBOs. Most KBOs have been found within
5\degr\ from the ecliptic, but this may constitute an observational
bias, since most searches have been (and are) performed close to the
ecliptic. Since the FSVS is mostly pointing away from the ecliptic, we
will be able to set limits on the inclination distribution of KBO's.  \\
{\bf $\bullet$ Solar Neighbourhood Objects}: 
The planned re-observations after one year will allow for the
detection of high proper-motion objects in the Solar
neighbourhood. These will be extremely important to constrain the
low-mass end of the IMF in the solar neighbourhood, to estimate the
relative contribution of the disk and halo population of stars in the
solar neighbourhood and trace the star formation history of the
galactic halo by finding old, high proper motion, white dwarfs. It may
also serve as a powerful tool to find nearby solitary field brown dwarfs.  

\section{The INT Wide Field Camera \label{sec:wfc}}  
The Wide Field Camera\footnote{see:
http://www.ast.cam.ac.uk/$\sim$wfcsur for an
extensive description of the WFC} 
(WFC) is mounted at the prime focus of the 2.5m
Isaac Newton Telescope (INT) on the island of La Palma. The WFC
consists of 4 EEV42 CCDs, each containing 2048$\times$4100 pixels. 
They are fitted in an
L-shaped pattern, which makes the Camera 6k$\times$6k, minus a
2k$\times$2k corner (see Figure 1). 
The CCDs consist of 13.5$\mu$ pixels (0\farcs33
per pixel on the sky), which gives a sky coverage per CCD of 
22\farcm8$\times$11\farcm4. A total of 0\fsqd29 is covered by the
combined four CCDs. 
With a typical seeing of 1\farcs0-1\farcs3 on the INT, point
objects are well-sampled, which allows for accurate photometry. The
Camera is equiped with a filter set consisting of Harris B,V and R,
RGO U, I and Z filters and Sloan
g$^{\prime}$,r$^{\prime}$,i$^{\prime}$ filters. Zeropoints,
defined as the magnitude that gives 1 detected e$^-$/s, of the
instrument are 25.6 in B,V and R, 23.7 in U and 25.0 in I.  

\begin{figure}
\centerline{\psfig{figure=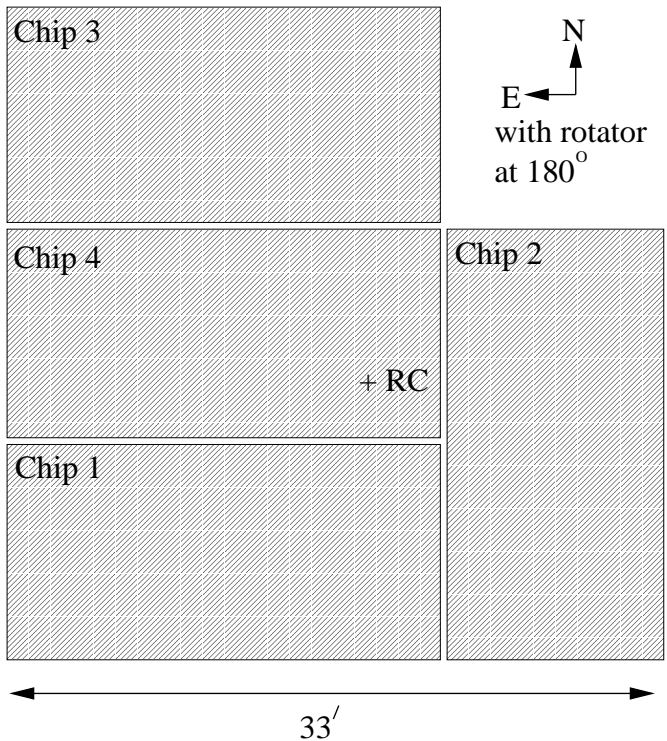,width=8.8cm,angle=-90}}
\caption[]{Graphical lay-out of the WFC 4 EEV 4k$\times$2k CCDs. In
the orientation used by the FSVS, North is up and East is to the
left. The WFC rotates around its Rotator Center (RC). \label{fig:layout}}
\end{figure}

\section{Observing strategy \label{sec:strategy}}

The typical timescales of variability covered by the objects listed
above vary from hours (CVs, KBOs, RR Lyraes) to days (Optical
transients to GRBs) to years (high proper motion stars). To cover all
possible timescales of variation we have devised an observing strategy
that optimises both the coverage per field as well as the total sky
coverage. The variability search is done in the V-band filter. This is
an optimum between the expected colours of our sources (blue as well
as red, sensitivity of the WFC (peaks in B and V) and the coverage of
the optical band with three filters, for which we have chosen the B
and I-band filters to obtain colour information.  For the photometric
variability we find that at least 15-20 pointings are needed to firmly
state that an object is variable and also get an indication of the
timescale of its variability (or ideally its period). For the first
two runs of the FSVS this number was limited to $\sim$12, but has been
raised to 15-20 in subsequent runs.  For the first two years the FSVS
has been allocated one week of dark time per semester. Since each
observing run consists of six to seven consecutive nights of dark
time, the 15-20 pointings per field have to be distributed over these
nights.  One of the criteria for the selection of the fields (see
Sect.\,5) is the fact that we want to observe each field within
30\degr\ of the zenith. Per night this gives an effective time-slot of
three hours to observe a particular field.  We have chosen to observe
the fields within 30\degr\ of the zenith to minimize the effect of
differential refraction on the differential photometry and astrometry
used to determine the light curve and proper motions of each object.

These considerations have led to a semi-logarithmic observing
sequence in which all possible time-scales are sample.  The exact
order of observations within the observing period depends on when a
photometric night occurs. It is on these nights that the photometric
calibration and field observations in B and I are done, along with two
observations in V. Integration times are 10 minutes in B and V and 15
minutes in I. Combined with the observing schedule outlined above,
this means that per three hour period three different fields can be
observed. In a six night run it is possible to observe two sets of
fields, that have an intertwining observing schedule.  In practice
this means that for observations in, for instance, November, when nights have
9-10 hours of dark time, we can observe 2 (sets of fields)$\times$3
(observing slots per night)$\times$3 (fields within one slot) = 18
fields in total. In November 1998 we were able to cover 18 fields
(=5\fsqd22). In May 1999 and May 2000, 
when observing nights were only 6-7 hours,
we covered 12 fields (3\fsqd48) each. After a year each field is
re-observed, enabling the search for long-term photometric variability
and high-proper motion objects.

\section{Field selection \label{sec:selection}}

The field selection is governed by the following four criteria (in
order of importance), which have been set to ensure maximum quality 
of the data:
\begin{itemize}
\item Fields are located between Galactic latitude 20\degr $< b^{II}
<$ 40\degr: to probe the Galactic halo as well as the Galactic disk to
considerable depths we target our fields at mid-Galactic
latitudes. This also prevents problems with field crowding and
interstellar extinction that will be present at lower Galactic
latitudes. The field crowding would limit the accuracy of the
differential photometry, especially for faint objects. The main effect
of interstellar extinction would be to limit the distance to which we
are able to observe into the halo.
\item Fields are observed within a zenith distance, $z<$30\degr:
this criterion has been set to ensure that the effect of differential
extinction coefficients has no impact on the accuracy with which the
differential photometry can be done.
\item If possible we will point our fields at the ecliptic, to
increase the chances of finding KBOs. However, as explained in Sect.\
\ref{sec:astgoals} even if we are not able to point at the ecliptic,
our results may help to constrain the inclination distribution of
KBOs.
\item Bright stars are avoided: stars brighter than $\sim$10th
magnitude will cause large charge overflows and diffraction patterns
that limit the area on a CCD that can be used for accurate
photometry, depending on the placement and brightness of the star. To
prevent this from happening the fields are selected to be as devoid as
possible of bright stars. We checked for the presence of
bright stars using the DSS in the selection of the fields. 
\end{itemize}
It is clear that not all four criteria can be met at all times of
the year. For the northern Hemisphere all four criteria are only
satisfied in late November-early December. 

\section{Comparison with other surveys \label{sec:comparison}}

The FSVS is unique in its search for variability on
short timescales (tens of minutes to days), depth and precision of its
differential photometry, although having a rather moderate sky-coverage. The
Sloan Digital Sky Survey (SDSS) covers a much larger area of the sky
(10\,000 \sqd), but at brighter magnitudes (14 $<g^{\prime}<$ 22.5),
and provides almost no variability information. The microlensing studies
(e.g. MACHO, Alcock et al., 1997; EROS, Beaulieu et al., 1995) 
do obtain variability information, but are targeted
at different stellar populations (the Galactic Bulge, the LMC, or M31)
and have a limit of V$\sim$21 with a photometric 
precision of 0.5 mag at the faint end, caused by
limited S/N and crowding in their necessarily high density star
fields. Supernovae searches reach as deep as the
FSVS, but have a much lower time-resolution.  In Figure 2 we
show schematically how the FSVS compares with other deep ongoing surveys.

\begin{figure}
\centerline{\psfig{figure=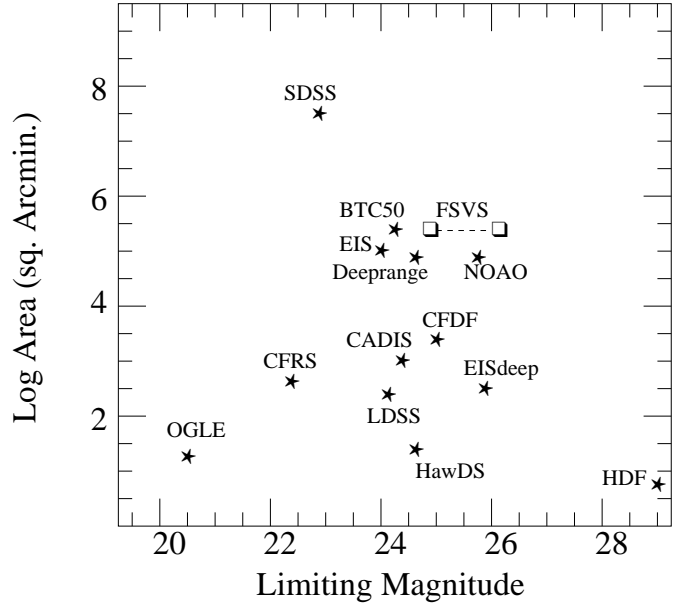,angle=-90,width=8.8cm}}
\caption[]{A comparison in area and depth between major current
surveys and the FSVS. Adapted from the NOAO Deep Survey
Web-pages (see http://www.noao.edu/noao/noaodeep/; SDSS=Sloan Digital
Sky Survey; BTC50 = Survey of Falco et al.; EIS(deep) = ESO
Imaging Survey (Deep); Deeprange = Postman et al I-band survey; HawDS= Hawking Deep Survey; HDF=
Hubble Deep Field; NOAO= NOAO Survey; CFRS = Canada France Redshift
Survey ; CADIS = Calar Alto Deep Imaging Survey; CFDF = Canadian French Deep Fields).
Note that most of these surveys have no or very limited variability
information. The range in depth for the FSVS is reached by using each
individual image (as in the variability study) or the sum images.}
 \end{figure}

\section{Reduction and Analysis Methods \label{sec:reduction}}

To obtain variability information on all the objects detected in our
observations we use the technique of differential aperture
photometry. We have written a pipe-line reduction package, consisting
of IRAF tasks, Fortran programs and at its core the SExtractor program
by Bertin and Arnouts (1996). 
Every object in every observation is analysed and the results
are stored in a master-table that lists the essential information
(described below in detail) for each object. When a photometric
calibration is available, the colours of each object are determined and
written to the master-table. Below we outline the data flow
through our pipe-line reduction, starting with the raw data as it
comes from the telescope. 

\subsection{Bias subtraction \label{sec:bias}} 

The mean of the counts in the overscan region of each observation is used to
subtract the overall bias level. After this the 2-D bias pattern,
determined from bias observations taken at the start of the night, is
subtracted. 

\addtocounter{footnote}{-1}
\subsection{Linearization of the data \label{sec:linear}}

A non-linearity in the read-out electronics causes all data taken with
the INT WFC to be non-linear up to a level of $\sim$5\%.  The
magnitude of this non-linearity as a function of exposure level is
determined by the Cambridge WFS group\footnote{ see:
previous footnote for URL} 
and is posted in tabular and analytic form. These corrections are applied
after bias-subtraction.  

\subsection{Flatfielding \label{sec:flatfield}}

From twilight skyflats taken during a whole observing run a
master flatfield is made, which is used for all the observations taken in
that band during the observing run. For the I-band observations, which
suffer from fringing at the 3.5\% level, we have made fringe maps from
the night time observations, which allows the fringe pattern to be removed
down to the 0.6\% continuum sky level (see Fig.\ \ref{fig:fringe}).

\begin{figure*}
\begin{minipage}{5cm}
\end{minipage}
\begin{minipage}{5cm}
\end{minipage}
\begin{minipage}{5cm}
\end{minipage}
\caption[]{Defringing of the I-band observations, using a fringe map
made from the night-time observations themselves. Left: Before
defringing, middle: fringe map, right: after defringing \label{fig:fringe}}
\end{figure*}

\subsection{Source detection \label{sec:detection}}

The bias-subtracted, linearized and flatfielded data are fed to
the SExtractor program. This
program detects sources and measures their instrumental magnitude in a
number of different ways, as set by the user. Source detection 
is done by requiring that three neighbouring pixels are more than
two sigma above the sky-background. Visual inspection shows that this
threshold value is capable of detecting virtually all objects that can be
identified by eye. Some contamination from extended cosmic rays is
present, but these are effectively removed in the subsequent
steps. Apart from finding the sources and determining their
instrumental magnitudes, 
for each source the SExtractor program determines various other
parameters such as its position, size, extent, ellipticity and
orientation angle.
Due to vignetting a corner of CCD3 (the NE corner in Fig.\
\ref{fig:layout}) has very low count rates. We discard any object
detected in a square box 200 pixels wide from this corner of CCD3. 

\subsection{Instrumental magnitudes \label{sec:instrumental}}

From bright, unsaturated objects detected in the central 1k$\times$1k
pixels of each CCD the seeing is determined from the median of the
distribution of 2-D Gaussian fits to those objects. This seeing
parameter is used to set the sizes of our photometry apertures for
each exposure. We measure the objects in four apertures having radii
of 0.5, 1.0, 1.5 and 2.0 times the seeing FWHM. These radii are chosen
to sample around the optimal S/N radios of 1.3$\times$the seeing
FWHM. However, depending on the object's brightness a smaller or
larger aperture than this optimal value may be applicable. We leave it
to the user to choose the aperture most suited for the topic under
investigation. Errors are calculated from the photon
counting statistics.

\begin{figure}
\centerline{\psfig{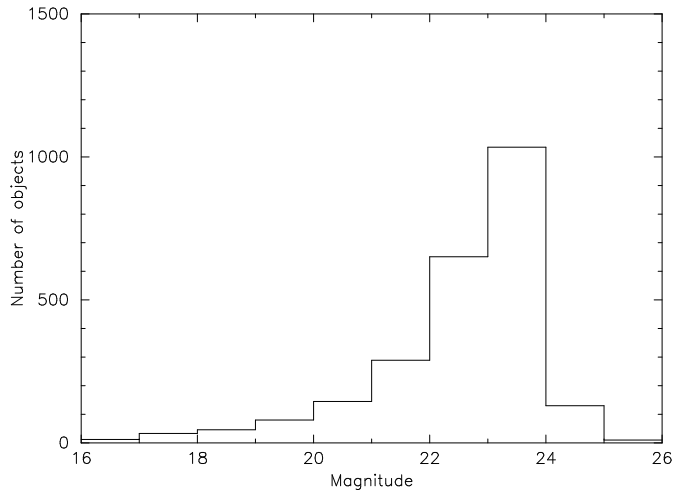}}
\caption[]{Histogram of the number of objects per magnitude for one
CCD in one of our lower galactic latitude fields. In total we detect $\sim$2300
objects with $V<$24. \label{fig:hist}}
\end{figure}

Using aperture photometry ofcourse relies on an accurate background
subtraction around each object. This fails when field crowding becomes
severe. However, even in the most crowded field (taken at a galactic
latitude, $b^{\sc ii}\sim$20\degr), the density of objects is still
low enough for good aperture photometry to be applied. We illustrate
this in Fig.\ \ref{fig:hist}, which shows the detection histogram for
one CCD in one of our lower galactic latitude fields. We detect
$\sim$2300 objects with $V<$24 in this image. With an average seeing
of 1\farcs2, each object occupies roughly 16 pix. In total these
objects will therefore cover about 3.7$\times$10$^3$ pix, a very small
fraction,$<0.5$\%, of the total area available on the CCD.

\subsection{Field matching \label{sec:matching}}

Different observations of the same field are automatically 
matched using the {\sc offset} program, supplied with the {\sc dophot}
package (Schechter, Mateo and Saha; 1993), using the 
100 brightest, non-saturated stars, that are not located near the
edges of the CCDs. Matching is done by
triangle pattern recognition in the two images. This matching allows for
linear scaling, rotation and translation of the different images. Output is
given as the elements of a rotation-translation matrix. All images are
transformed to one of the images that is taken as a reference image
(typically the one with the best seeing). Individual objects are
matched if in the new image an object is found within 1 FWHM of the
position of the object in the reference image. Given the low density
of objects in our images the chances of an incorrect matching are very
low. This same criterion is used to match stars between different
filters.  

\subsection{Local reference star selection \label{sec:standards}}

In order to obtain differential magnitudes, an ensemble of local
reference stars has to be selected. The average (ensemble) magnitude
of these stars is used to compute all instrumental magnitudes. In the
selection of this ensemble it is important to use the brightest,
non-variable, stars that are not saturated. Using the brightest stars
is essential because the error on the differential magnitude of any
object consists of the error that is obtained from counting statistics
for that object, and the error on the average of the reference stars
(see e.g. Howell, Mitchell and Warnock, 1988). The uncertainty in the
mean magnitude of the ensemble must be made significantly smaller than
the uncertainty imposed by counting statistics on the magnitude of any
star of interest.  If this is not the case, it will cause
small-amplitude variability, that should have been detected on the
basis of counting statistics, to become undetectable. Currently, per
CCD, an ensemble of at least ten local standards is selected by
requiring that their variation with respect to the average is less
than 5 millimagnitudes. If this requirement is set more stringently not
enough standards are found. In the North Galactic Pole observations of
May 1999 the selection criterion had to be relaxed to 10
millimagnitudes in order to find a suitable number of stars. This is,
of course, due to the limited number of stars in the NGP direction. As
explained above, this selection criterion naturally sets the minimum
amplitude of variation that can be found.

\subsection{Differential magnitudes \label{sec:difmags}}

For every object the differential magnitude is calculated against the
ensemble average. The error of the instrumental magnitude is
propagated to the differential magnitude, adding quadratically the
error on the ensemble average. The differential magnitude is
calculated for all four aperture size as described in Sect.\
\ref{sec:instrumental}. 

\subsection{Absolute calibration \label{sec:abscal}}

Using the USNO A2.0 catalogue an astrometric solution is
obtained for each CCD and each field separately. 
On average, each CCD contains 20-30 USNO A2.0 stars, which 
is sufficient to obtain a cubic solution that is accurate to
0\farcs2-0\farcs4 in right ascension and declination, depending on the
position of a field on the sky.

During all our runs so far, we have had two photometric nights, during
which all fields and several Selected Areas of Landolt (1992) were
observed. After having found the astrometric solution, we can measure
the standard stars automatically, now knowing where they are located,
using their position in the Landolt catalogue. We use the SExtractor
aperture photometry option, with an aperture radius of twice the image
FWHM. For each CCD the measured BVI standard star magnitudes are fit
with a model that includes a zero-point offset, an airmass term and a
colour term. When sufficient standards are observed at different
airmasses, we fit for the airmass term. If not, we hold it constant at
the following values: 0.25, 0.15 and 0.07 for the filters B, V and I,
respectively\footnote{see http://www.ast.cam.ac.uk/$\sim$wfcsur}.  The
colour term is only included if it improves the fit significantly.
These solutions are applied to all objects listed in the catalogue
through the reference stars that are selected for each CCD of each
field (see Sect. 7.7). We estimate the error in the absolute
calibration to be 0.05 for the B and V filters, and 0.1 for the I
band.

\subsection{Limiting magnitudes \label{sec:limmag}}

Based on the amount of flux in the ten
reference stars (see Sect.\ \ref{sec:standards}), the level of the
background sky, the photometry aperture size and the background
aperture size, we calculate the flux a 3-, 5-, and 7-sigma object would
have for each CCD, field and observation. 
These ten estimates of the 3-,5- and 7-sigma limits are then averaged
to produce an average 3-, 5- and 7-sigma limiting magnitude.
In this calculation we neglect the read-out noise since our
observations are long and have background levels whose noise
is much higher than the read-out noise. 



\subsection{Variability \label{sec:var}}

Variability of an object is determined on the basis of the differential
magnitudes discussed above. The mean of the differential
light curve of each object. 
This constant fit returns a $\chi^2$-value, the magnitude
of which is taken as a measure of the object's variability. In case
an object is not detected in an observation, that observation's
limiting magnitude is taken as an upper limit to the brightness of the
object. 

\subsection{Star- Galaxy seperation}

\begin{figure}
\centerline{\psfig{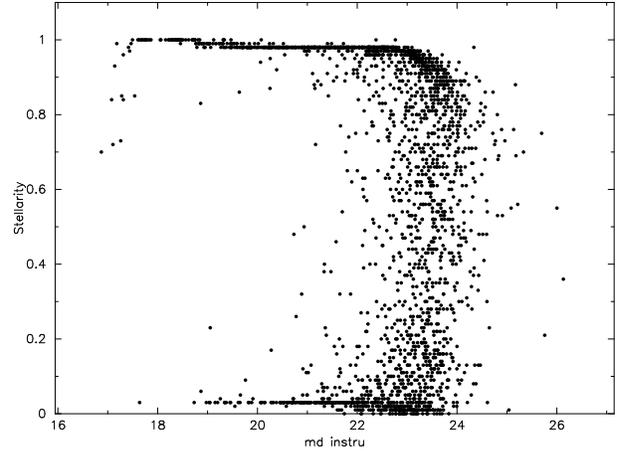}}
\caption[]{The stellarity versus magnitude for one CCD in one of our
fields. A stellarity of zero indicates a highly extended source, and a
stellarity of one is a point-source. Different symbols represent
different observations. The lower stellarity of point-sources at the
bright end of our magnitude range is caused by saturation.
Detections at V$>$25 are noise spikes. 
\label{fig:stellarity}}
\end{figure}

The star-galaxy separation used in the FSVS is based on the
'stellarity' parameter, as returned from the SExtractor routines
(Bertin and Arnouts, 1996).  This parameter has a value between 0
(highly extended) and 1 (point source). In the FSVS the stellarity 
value of an object is taken as the value in the combined V-band
images. Due to the increased S/N is this image, the star-galaxy
separation can be done reliably almost 1 magnitude deeper than from 
any individual image. As can be seen in Fig.\
\ref{fig:stellarity} this seperation of object types works very well
to classify stars (with a value $>$0.8) down to V$\sim$23.5-24).
Fainter stars tend to have slightly lower stellarity values (the turn
down between 23 and 24) but can still be well separated from the
galaxies. 

\subsection{Astrometric data-analysis and search for Kuiper Belt
Objects}

The search for Kuiper Belt Objects is made using the moving object
detection code of one of us (HS).  This code is tuned to the particular
photometric properties of the data frames in question as well as to the
expected range of rate of motion for KBOs in that area of the sky.
The signal-to-noise cutoff is selected to find the maximum number of
real objects while keeping the rate of false detections to a manageable
level.  
 
Three, or ideally four images, taken at different times during the same night
are scanned by the Scholl code to identify all "stellar" objects on
the frames and match them across frames.  Those with rates of motion
within the target range are flagged and small cutouts of the CCD frames
including each of the flagged objects are automatically prepared.  These
are blinked by a human operator to confirm the detections of real
objects and remove false detections.

Once the real detections have been flagged, the code performs
astrometry on the objects using the USNO catalogue.  
The resulting astrometry is ready for reporting to the Minor
Planet Center and for follow-up observations.

\section{Final products \label{sec:products}}

The pipeline discussed above returns two sets of output files:\\
$\bullet$ The reduced images\\
$\bullet$ The data tables with the photometric and astrometric
information.\\
The data tables are made per field, per CCD and are made for four
different apertures: 0.5, 1.0, 1.5 and 2.0 $\times$FWHM, where the
FWHM is defined as the average full width at half maximum of the
stellar profiles in the central 1k$\times$1k part of each CCD and
each exposure. 

The data tables contain, for all the detected objects,
the information on the time of observation, name, position and
colour for each object, followed by the magnitude, error on the
magnitude, fwhm, stellarity and the error flag as returned from the
SExtractor program for each object and each observation. 

If an object is only detected in a subset of all the observations, it
is added to the final catalogue, and dummy values are introduced when
it was not detected. 

The objects names are given in standard IAU format as FSVSJhhmmss.ss+ddmmss.s,
all in J2000 coordinates. Each object is also given an 'internal' name
whose format is F\_XX\_Y\_ZZZZZ, with XX the field number, Y the CCD
number (1-4) and ZZZZZ a five digit detection number. The position of
each object is given both in RA and DEC as well as in x,y-coordinates
in the reference frame of the specific field. 
  

\section{Availability of the data \label{sec:data}}

All raw data is available upon request from the ING-WFS archive in
Cambridge after the one year propietary right. For UK and NL
astronomers the data is immediately available.
All data-tables, containing the reduced information described
above, are retrievable from the
FSVS-website\footnote{http://www.astro.uva.nl/$\sim$fsvs}.  

\section{First year observations \label{sec:status}}

An extensive discussion of the results from the first year of observations
is given in Paper II and is outside the scope of this
paper. However, observing conditions in the first two runs of the
FSVS have been good and data is now available for a total of 30 fields
(8\fsqd7). As will be
discussed in Paper II, using the pipe-line reduction as described
above, point source light curves have been obtained with photometric
precisions as good 
as a few millimag for the brightest (V$\sim$17) sources. At 24th
magnitude precision on the differential photometry is still in the
0.1-0.15 mag range. 

\section{Conclusions}

The FSVS offers a unique possibility of studying the behaviour
of variable objects in the magnitude range of 17 $< V <$ 25 with
photometric precisions ranging from 3 millimag (at $V$=17) to 0.15 mag (at
V$\simeq$24). Observations in the first year show that the FSVS is
producing promising results. 

Besides the study of variable objects, the FSVS offers a large dataset
that can serve as the basis for many research topics (e.g. YSO's,
gravitational lenses, galaxy counts, quasar searches). The
FSVS-collaboration encourages the use of the data set for purposes
other than the ones mentioned here.


\section*{acknowledgement}

PJG, PMV and the Faint Sky Variability Survey are supported by NWO
Spinoza grant 08-0 to E.P.J.van den Heuvel. PJG is also supported by a CfA
fellowship. SBH acknowledges partial support of this research from NSF
grant AST 98-19770. MEH is partially supported by a NASA/Space Grant
Fellowship, NASA Grant \#NGT-40008. The FSVS is part of the INT Wide
Field Survey. The INT is operated on the island
of La Palma by the Isaac Newton Group in the Spanish Observatorio del
Roque de los Muchachos of the Instituto de Astrofisica de Canarias

\end{document}